\documentclass[12pt]{article}
\pdfoutput=1
\usepackage[latin1]{inputenc}

\usepackage{amsmath}
\usepackage{amsfonts}
\usepackage{amssymb}
\usepackage{graphicx}
\usepackage{geometry}
\usepackage{amssymb,epsfig}
\usepackage{hyperref}
\usepackage{subfig}

\makeatletter
\renewcommand\section{\@startsection {section}{1}{\z@}%
                                 {-3.5ex \@plus -1ex \@minus -.2ex}
                                   {2.3ex \@plus.2ex}%
                                   {\normalfont\large\bfseries}}
\renewcommand\subsection{\@startsection{subsection}{2}{\z@}%
                                   {-3.25ex\@plus -1ex \@minus -.2ex}%
                                     {1.5ex \@plus .2ex}%
                                     {\normalfont\bfseries}}
\renewcommand\subsubsection{\@startsection{subsubsection}{3}{\z@}%
                                   {-3.25ex\@plus -1ex \@minus -.2ex}%
                                     {1.5ex \@plus .2ex}%
                                     {\normalfont\itshape}}
\makeatother

\def\pplogo{\vbox{\kern-\headheight\kern -29pt
\halign{##&##\hfil\cr&{\ppnumber}\cr\rule{0pt}{2.5ex}&\ppdate\cr}}}
\makeatletter
\def\ps@firstpage{\ps@empty \def\@oddhead{\hss\pplogo}%
  \let\@evenhead\@oddhead 
}
\def\maketitle{\par
 \begingroup
 \def\thefootnote{\fnsymbol{footnote}}
 \def\@makefnmark{\hbox{$^{\@thefnmark}$\hss}}
 \if@twocolumn
 \twocolumn[\@maketitle]
 \else \newpage
 \global\@topnum\z@ \@maketitle \fi\thispagestyle{firstpage}\@thanks
 \endgroup
 \setcounter{footnote}{0}
 \let\maketitle\relax
 \let\@maketitle\relax
 \gdef\@thanks{}\gdef\@author{}\gdef\@title{}\let\thanks\relax}
\makeatother

\numberwithin{equation}{section}

\renewcommand{\dag}{\dagger}

\newcommand{\be}{\begin{equation}}
\newcommand{\bea}{\begin{eqnarray}}
\newcommand{\ee}{\end{equation}}
\newcommand{\eea}{\end{eqnarray}}
\newcommand\beq{\begin{equation}}
\newcommand\eeq{\end{equation}}

\newcommand{\mc}{\mathcal}

\newcommand{\tr}{{\rm tr}}
\renewcommand{\t}{\tilde}

\newcommand{\ba}{\begin{align}}
\newcommand{\ea}{\end{align}}
\newcommand{\bg}{\begin{gather}}
\newcommand{\eg}{\end{gather}}
\newcommand{\bseq}{\begin{subequations}}
\newcommand{\eseq}{\end{subequations}}

\newcommand{\eneq}{\end{equation}}
\newcommand{\ff}{\varphi}
\newcommand{\reff}[1]{(\ref{#1})}
\newcommand{\IR}{\overset{g\rightarrow\infty}{=}}

\begin{document}

\setcounter{page}0
\def\ppnumber{\vbox{\baselineskip14pt
}}
\def\ppdate{\footnotesize{SLAC-PUB-15461, SU-ITP-13/05}} \date{}

\author{Fabio Saracco$^1$, Alessandro Tomasiello$^1$, Gonzalo Torroba$^2$\\
[7mm]
{\normalsize $^1$\it Dipartimento di Fisica, Universit\`a di Milano-Bicocca, I-20126 Milano, Italy, and}\\
{\normalsize  \it INFN, sezione di Milano-Bicocca, I-20126 Milano, Italy}\\
\\
{\normalsize $^2$\it Stanford Institute for Theoretical Physics and SLAC}\\
{\normalsize \it Stanford University, Stanford, CA 94305, USA}
}

\bigskip
\title{\bf Topological resolution of gauge theory singularities
\vskip 0.5cm}
\maketitle

\begin{abstract}
Some gauge theories with Coulomb branches exhibit singularities in perturbation theory, which are usually resolved by nonperturbative physics. In string theory this corresponds to the resolution of timelike singularities near the core of orientifold planes by effects from F or M theory. We propose a new mechanism for resolving Coulomb branch singularities in three dimensional gauge theories, based on Chern--Simons interactions. This is illustrated in a supersymmetric $SU(2)$ Yang--Mills--Chern--Simons theory.
We calculate the one loop corrections to the Coulomb branch of this theory and find a result that interpolates smoothly between the high energy metric (that would exhibit the singularity) and a regular singularity-free low energy result. We suggest possible applications to singularity resolution in string theory and speculate a relationship to a similar phenomenon for the orientifold six-plane in massive IIA supergravity.
\end{abstract}
\bigskip
\newpage

\tableofcontents

\vskip 1cm

\noindent

\vspace{0.5cm}  \hrule
\def\thefootnote{\arabic{footnote}}
\setcounter{footnote}{0}

\section{Introduction and summary}\label{sec:intro}

Understanding spacetime singularities is one of the central problems in string theory. Singularities arise in diverse systems, ranging from cosmology to black holes. So far we don't know the necessary and sufficient conditions for a singularity to be resolved\footnote{Some proposals in the context of AdS/CFT were given in~\cite{Gubser:2000nd,Maldacena:2000mw}. }, and lacking a complete framework to answer this question it is important to identify concrete mechanisms that can resolve singularities. In this work we will present a novel mechanism for resolving certain timelike singularities, based on Chern--Simons interactions.

One successful approach has been to exploit the connection between D-branes and geometry. Using D-branes to probe string theory solutions reformulates the spacetime geometry in terms of the scalars on the worldvolume gauge theory. This is helpful because the gauge theory can capture phenomena that are beyond the reach of the classical gravity description.\footnote{For more details and references see~\cite{Polchinski:1998rr,Johnson:2003gi}.} Here we will focus on gauge theories with perturbative Coulomb branch singularities. These describe certain timelike singularities of string theory, the canonical example being that of orientifold planes. 

From the gauge theory side, one of the first and most important examples of a Coulomb branch singularity appeared in the work of Seiberg and Witten~\cite{Seiberg:1994rs} on the $SU(2)$ Yang--Mills theory (YM) with $\mc N=2$ supersymmetry in four dimensions. The gauge coupling along the Coulomb branch receives a one-loop correction
\be\label{eq:4dsing}
\frac{1}{g^2(\phi)} = \frac{1}{g_0^2}+ \frac{1}{4\pi^2} \log \frac{|\phi|^2}{\Lambda^2}\,,
\ee
where $\phi$ is the Coulomb branch coordinate, $g_0$ is the UV value of the gauge coupling, and $\Lambda$ is the dynamical scale. This result, which is exact in perturbation theory, shows a singularity at $|\phi|^2 = \Lambda^2 e^{-4\pi^2/g_0^2}$. For smaller values of $|\phi|^2$ the perturbative answer cannot be correct because it predicts a negative gauge coupling squared. Since the QFT is UV complete, it should make sense at arbitrarily long distances; so the failure of perturbation theory suggests that new physics should become important at the scale $\Lambda^2 e^{-4\pi^2/g_0^2}$. And indeed, Seiberg and Witten showed how this singularity is resolved by nonperturbative effects, and found that the correct low energy description is in terms of a light monopole and a dyon near the origin of the moduli space.

Another example, which will be relevant for us, is obtained by compactifying one spatial dimension in this theory. At low energies this gives an $SU(2)$ Yang--Mills theory with $\mc N=4$ supersymmetry in three dimensions (i.e. 8 supercharges). Now the gauge coupling including perturbative corrections is~\cite{Seiberg:1996nz, Dorey:1997ij}
\be\label{eq:3dsing1}
\frac{1}{g^2(\phi)} = \frac{1}{g_0^2}- \frac{1}{\sqrt{8}\pi |\phi|} \,.
\ee
Nonperturbative contributions from instantons resolve the perturbative singularity at $|\phi| =g_0^2/(\sqrt{8} \pi)$, giving rise to the smooth Atiyah--Hitchin metric~\cite{Atiyah:1988jp}.

In these examples, the existence of singularities signals the appearance of nontrivial nonperturbative dynamics. Our goal is to understand what happens to such singularities when these nonperturbative effects are absent. Let us deform the previous theory by adding a topological Chern--Simons (CS) mass that does not lift the Coulomb branch. Intuitively, the topological mass for the gauge field leads to confinement of monopole-instantons~\cite{Pisarski:1986gr}, and we will argue that the Yang-Mills instantons deformed by a CS term do not have finite action. Therefore, they cannot resolve the moduli space singularity. How is then the Coulomb branch singularity resolved?

In this work we analyze the quantum corrections to the Coulomb branch of Yang--Mills--Chern--Simons theories and show that the Chern--Simons interaction resolves the singularity (\ref{eq:3dsing1}) already in perturbation theory, without the need of nonperturbative contributions. We focus on the simplest theory that has a Coulomb branch which receives nontrivial loop corrections; this is the previous $SU(2)$ gauge theory plus a CS term that preserves $\mc N=2$ supersymmetry. Besides our motivation from singularities, this class of theories may be relevant to the AdS/CFT correspondence for 3d gauge theories, or the intriguing possibility of emergent supersymmetry in condensed matter systems~\cite{Grover:2012bm}. Furthermore, as far as we are aware, there is not much work on the quantum Coulomb branch of YM--CS theories, and we hope that our results help to bridge this gap.

Before proceeding to the explicit analysis, let us discuss the relation of this mechanism to timelike singularities in string theory. One of our motivations was to understand the gauge theory version of~\cite{Saracco:2012wc}, who found that the O6 plane singularity is resolved in massive type IIA compactifications to four dimensions. In the absence of Romans mass, the O6 plane geometry can be understood using a D2 probe, which gives exactly (\ref{eq:3dsing1}). The strong coupling limit of type IIA is M-theory, and the nonperturbative effects discussed before arise from exchange of D0 branes~\cite{Polchinski:1997pz}. However, it was argued in~\cite{Aharony:2010af}  that massive type IIA does not have an M-theory limit. Instead, the O6 singularity is resolved by the Romans mass~\cite{Saracco:2012wc}. A D2 probe in massive IIA acquires a Chern--Simons term, and this is the basic effect that our mechanism captures.

Actually, it turns out that there exists another type of orientifold plane, the $\widetilde{O6}$, whose singularity resolution is not yet known. This plane can be thought of as an O6 plus a half D6 brane~\cite{Hyakutake:2000mr}. The half D6 introduces a fundamental flavor on the D2 probe which produces, at one loop, a Chern--Simons mass for the worldvolume gauge field. Thus, we expect that our results will be relevant to understanding the $\widetilde{O6}$. Another possible application would be to consider Chern--Simons deformations of the enhan\c{c}on mechanism of~\cite{Johnson:1999qt} for the resolution of repulson-type singularities.

Here we will focus on the previous three dimensional Yang--Mills--Chern--Simons theory with $\mc N=2$ supersymmetry, postponing to a future work a more detailed exploration of the effects on timelike singularities. One reason for this is that, with this amount of supersymmetry, the Coulomb branch metric is not protected against higher loop corrections. Our results, valid at one loop and in the perturbative limit, can receive nontrivial corrections in the regime of interest for the gravitational singularities. Also, in string theory solutions such as that of~\cite{Saracco:2012wc}, we expect the probe theory to be more complicated, with reduced supersymmetry and only an approximate moduli space. The theory that we will study is the simplest one where our mechanism can be exhibited under controlled approximations. Nevertheless, we expect that our conclusions are applicable more generally. \\

The paper is organized as follows. First, in \S \ref{sec:YMCS} we introduce the Yang--Mills--Chern--Simons theory, study the classic Coulomb branch and show that the YM instantons have infinite action in the presence of the CS deformation. The rest of the work is then devoted to understanding the quantum geometry of the Coulomb branch in perturbation theory.

Our analysis of the quantum Coulomb branch metric is done in three steps. In \S \ref{sec:UVYM} we discuss the UV limit where the Coulomb branch expectation value $|\phi| \gg k g$, where $g$ is the gauge coupling and $k$ is the CS level. In this regime the CS deformation can be neglected, and we perform an explicit one loop calculation that reproduces (\ref{eq:3dsing1}). A nonzero expectation value for the Coulomb branch scalar $\phi$ leads to massive states of mass $m_g^2 = 2 g^2 |\phi|^2$, and the quantum corrections to the low energy theory are obtained by integrating out these fields. 

Next, in \S\ref{sec:CS} we study the quantum Coulomb branch in the CS theory ignoring the YM term, which corresponds to the IR limit  $g \to \infty$ at fixed $\phi$ and $k$ --- up to important finite effects that we discuss below. We present a general argument explaining why the CS interaction resolves the singularity: for dimensional reasons, scalar expectation values cannot appear in the corrections to the metric, which thus has to be finite. We also present an explicit calculation showing indeed a finite smooth correction to the Coulomb branch metric of order $1/k$. 

Finally, in \S \ref{sec:fullYMCS} we determine the quantum Coulomb branch metric in the full YM--CS theory at one loop. 
The effect of the CS deformation on the massive fields is to split $m_g^2 \to m_{\pm}^2$, where the mass eigenvalues $m_{\pm}^2$ are defined in (\ref{eq:mpm}). Surprisingly, the one loop quantum corrected Coulomb branch metric is exactly the same as in the pure YM case, after replacing the Higgs mass $m_g$ by the average mass of the massive modes in the presence of the CS deformation, $(m_+ + m_-)/2$. This gives the Coulomb branch metric
\be 
S \supset \int d^3 x \left(1- \frac{1}{\sqrt{8\pi^2 \frac{|\phi_3|^2}{g^2}+ \frac{k^2}{16}}} \right) \partial_\mu \phi_3 \partial^\mu \bar \phi_3\,.
\ee
This result interpolates smoothly between (\ref{eq:3dsing1}) in the UV and a finite nonsingular metric in the IR that becomes independent of $\phi$. 
Furthermore, the answer in the deep IR disagrees with the one calculated in the pure CS theory in \S \ref{sec:CS} by a factor of $2$. This is due to an interesting nondecoupling effect from fields that become infinitely massive in the limit $g \to \infty$ but still contribute finite corrections. This will end our analysis, proving our claim that the CS interaction resolves the moduli space singularity by perturbative quantum corrections. Some useful formulas that are needed for our analysis are collected in the Appendix.

\section{The Yang--Mills--Chern--Simons theory}\label{sec:YMCS}

Let us begin by describing the classical Yang--Mills--Chern--Simons (YM--CS) theory that will be the subject of our analysis.
 This is a three dimensional $SU(2)$ gauge theory with $\mc N=2$ supersymmetry --- 	i.e. four supercharges. The matter content consists of a vector superfield with components $(A_\mu^a, \sigma^a, \lambda^a, D^a)$ and a chiral superfield $(\phi_a, \psi_a, F_a)$. Here $a$ is the index for the adjoint representation of the gauge group, $\sigma$ is a real scalar\footnote{This scalar is the extra component of the gauge field upon compactifying the 4d theory on a circle.}, $\lambda$ and $\psi$ are 3d Dirac fermions, and $D$ and $F$ are auxiliary fields. The theory contains a CS deformation for the vector supermultiplet and there is no superpotential. A review of three-dimensional supersymmetric YM--CS theories is given in the Appendix.
 
The general lagrangian with $\mc N=2$ supersymmetry and no superpotential is (see (\ref{eq:LCSfinal}))
\be\label{eq:L1}
L = L_{gauge}+ L_{CS}+L_{matter}
\ee
where the gauge kinetic terms are
\be
L_{gauge}=-\frac{1}{4} F_{\mu\nu}^a F^{a\mu\nu}+ \frac{1}{2} (D_\mu \sigma^a)^2 + i \bar \lambda^a \not\! \! D \lambda^a -g\bar \lambda \sigma \lambda\,,
\ee
the CS deformation is
\be
L_{CS}=\frac{1}{2} m_k \left[\epsilon^{\mu \nu \rho} (A_\mu^a \partial_\nu A_\rho^a + \frac{1}{3}g f^{abc} A_\mu^a A_\nu^b A_\rho^c) -(\sigma^a)^2-2 \bar \lambda^a \lambda^a  \right]\,,
\ee
and the remaining matter contributions and interactions are
\bea
L_{matter}&=& (D_\mu \phi_i)^\dag (D^\mu \phi_i)+ i \bar \psi_i \not  \!  \! D  \psi_i - g^2 \phi^\dag \sigma^2 \phi -g \bar \psi \sigma \psi + \sqrt{2} i g(\phi^\dag \bar \lambda \psi - \bar \psi \lambda \phi)\nonumber\\
&-&\frac{1}{2}\left( g \phi_i^\dag T^a_{ij} \phi_j\right)^2 - m_k g \sigma^a\phi_i^\dag T^a_{ij} \phi_j\,.
\eea
The auxiliary fields have already been integrated out, and we have introduced the mass parameter
\be
m_k \equiv \frac{g^2 k}{4\pi}
\ee
where $k$ is the CS level. Note that the CS deformation does not give a mass to the chiral supermultiplet.

Our conventions for covariant derivatives, index contractions, etc., are described in detail in the Appendix. Here we will specialize to an $SU(2)$ gauge group, with the chiral superfield also transforming in the adjoint representation. In this case, the indices $i,j$ also run over $1, 2, 3$, and $T^a_{ij}=-i \epsilon^{aij}$.

In the limit $m_k \to 0$ supersymmetry is enhanced to $\mc N=4$. This is the low energy limit of the four-dimensional pure $SU(2)$ theory with 8 supercharges compactified on a circle. This theory was studied by~\cite{Seiberg:1996nz}. Here it arises as the UV limit of (\ref{eq:L1}), while the IR is dominated by the CS deformation. These limits will be studied in \S\S \ref{sec:UVYM} and \ref{sec:CS} respectively, before tackling the whole problem.

\subsection{The classical Coulomb branch}\label{subsec:classicCoulomb}

The classical Coulomb branch of the theory is parametrized by the gauge invariant ${\rm tr}\,\phi^2$. Without loss of generality, we will choose a real expectation value along the 3rd direction in color space,
\be\label{eq:vev}
\langle \phi_a \rangle = v \delta_{a3}\;,\qquad v \in \mathbb R\,,
\ee
which breaks $SU(2) \to U(1)$. We will now determine the low energy description of this $U(1)$ theory, valid at energies $E \ll v$. 

We denote the color indices perpendicular to the Coulomb branch direction by $\alpha=1,2$, and split the scalar fields into real and imaginary parts
\be\label{eq:chidef}
\phi_\alpha = \frac{\chi_\alpha+ i \t \chi_\alpha}{\sqrt{2}}\ ,\qquad\phi_3 = v +\frac{\chi_3+ i \t \chi_3}{\sqrt{2}}\,.
\ee
The massive fields come from $(A^\mu_\alpha, \sigma_\alpha, \phi_\alpha, \lambda_\alpha, \psi_\alpha)$, with $\alpha=1,2$. The nonzero expectation value $v$ introduces another mass parameter
\be\label{eq:mgdef}
m_g^2 \equiv 2 g^2 v^2\,.
\ee
(Recall that both $\phi$ and the gauge coupling have classical dimension $1/2$).

Expanding (\ref{eq:L1}) to quadratic order around (\ref{eq:vev}) obtains,
\bea\label{eq:Lquad}
L_{quad}&=& \frac{1}{2} A_\alpha^\mu \left[g_{\mu\nu}(\Box + m_g^2) - \partial_\mu \partial_\nu - m_k \epsilon_{\mu \nu \rho} \partial^\rho \right] A_\alpha ^\nu - \frac{1}{2} \chi_\alpha \Box \chi_\alpha -\frac{1}{2} \sigma_\alpha (\Box + m_g^2+m_k^2) \sigma_\alpha \nonumber\\
&-&\frac{1}{2} \t \chi_\alpha (\Box + m_g^2) \t \chi_\alpha + m_g m_k \,\epsilon_{\alpha \beta}\, \sigma_\alpha \t \chi_\beta +m_g \,\epsilon_{\alpha \beta}\, \partial_\mu A^\mu_\alpha \,\chi_\beta \\
&+& \bar \lambda_\alpha (i \not  \!\partial - m_k) \lambda_\alpha + i \bar \psi_\alpha \not \! \partial \psi_\alpha - \epsilon_{\alpha \beta} m_g (\bar \lambda_\alpha \psi_\beta+ \lambda_\alpha \bar \psi_\beta)\nonumber\,.
\eea
The Lagrangian for the fields along $a=3$ is not modified by the expectation value and can be read off directly from (\ref{eq:L1}), so we have not included those terms here.
In the limit when the gauge symmetry becomes global, the fields $\chi_\alpha$ are the Goldstone bosons of the broken symmetry; that is why they do not have a mass term at this stage, and they couple derivatively to the massive vector bosons. This is, however, gauge dependent, and we will add a gauge fixing term below. We will first work in Landau gauge, and then consider an arbitrary $R_\xi$ gauge that will allow us to prove the gauge invariance of our results.

Let us now discuss the masses of the heavy fields. Since the bosonic part of the previous Lagrangian will be modified by the gauge fixing term, we will instead focus on the fermionic masses, which will not be altered. Diagonalizing the mass matrix obtains the mass eigenvalues
\be\label{eq:mpm}
m_{\pm}^2 \equiv m_g^2 + \frac{1}{2} m_k^2 \pm \frac{1}{2} m_k \sqrt{4m_g^2 + m_k^2}\,.
\ee
In the UV limit $m_g^2 \gg m_k^2$, $m_{\pm}^2 \sim m_g^2$, corresponding to both the gaugino and $\psi$ fermion acquiring the same mass (and consistent with the approximate $\mc N=4$). On the other hand, approaching the origin of the Coulomb branch $m_g^2 \ll m_k^2$, the $\psi$ fermion becomes massless, while the gaugino acquires a mass $m_k$. This is the topological mass induced by the $\mc N=2$ CS deformation.

In our calculation of quantum effects we will also need the interactions for the fields along the color directions $\alpha$. The interactions that contribute at one loop order are
\bea\label{eq:Lint1}
L_{int} &=&  -g\left( m_g  \chi_3 + \frac{1}{2} g \chi_3^2+\frac{1}{2}g \t \chi_3^2\right)\left(-(A_\mu^\alpha)^2 + \sigma_\alpha^2 + \t \chi_\alpha^2\right)+ g (m_g + g \chi_3) \chi_\alpha \t \chi_\alpha \t \chi_3 + \nonumber\\
&-&g m_k \epsilon_{\alpha \beta} \sigma_\alpha (\t \chi_\beta \chi_3 - \chi_\beta \t \chi_3)+ 2g \epsilon_{\alpha \beta} A_\mu^\alpha  (\chi_\beta \partial_\mu \chi_3 + \t \chi_\beta \partial_\mu \t \chi_3) +g \epsilon_{\alpha \beta}\partial_\mu A_\alpha^\mu  (\chi_\beta \chi_3 + \t \chi_\beta  \t \chi_3)  \nonumber\\
&-&g \epsilon_{\alpha \beta} \chi_3 (\bar \lambda_\alpha \psi_\beta+ \lambda_\alpha \bar \psi_\beta) + i g \epsilon_{\alpha \beta} \t \chi_3(\bar \lambda_\alpha \psi_\beta- \lambda_\alpha \bar \psi_\beta)\,.
\eea

\subsection{Analysis of instanton solutions }\label{subsec:instantons}

Now we come to the important question of instanton solutions in the YM--CS theory. The gauge theory in the absence of a CS deformation has three dimensional instanton solutions that resolve the perturbative Coulomb branch singularity. These instantons are the same as the 4d BPS monopoles of the Seiberg-Witten theory and are described for instance in~\cite{Dorey:1997ij}. We would like to understand what happens to these solutions when the theory is deformed by a CS term. Instantons in a nonsupersymmetric theory with a massive Higgs fields have been studied by several authors; see for instance~\cite{Pisarski:1986gr,Affleck:1989qf,Fradkin:1990xy,Tekin:1998qy}. Here we will adapt some of these techniques to the supersymmetric case, and argue that the real instantons of the YM theory have infinite action once the effects of the CS interaction are taken into account.\footnote{We do not claim that there are no finite action classical solutions; for instance, \cite{Tekin:1998qy} argue that there are complex solutions that can contribute to the path integral. The role of these solutions in the physical theory is not fully understood yet. However, these subtleties will not affect our main conclusion, namely the smoothness of the quantum corrected Coulomb branch.}

Choosing a Coulomb branch branch coordinate along the real part of $\phi_a$ as in (\ref{eq:vev}), the YM instantons have a nontrivial profile~\cite{Dorey:1997ij}
\begin{equation}\label{eq:Polyakov}
A_\mu^a\sim\epsilon_{a\mu\rho}\frac{x^\rho}{x^2}\;,\qquad \varphi^a \equiv \sqrt{2} \,{\rm Re} \,\phi^a \sim \frac{x^a}{x}\,,
\end{equation}
where $x$ is the euclidean distance.
The other bosonic fields are set to zero, and the solution for fermions can be obtained by a supersymmetry transformation.

Let us now add the CS term and find how (\ref{eq:Polyakov}) is modified. Following~\cite{Pisarski:1986gr}, we look for deformed instantons that are invariant under the diagonal subgroup of the rotations and gauge transformations groups:
\be\label{eq:PisAns}
A_\mu^a(x)=\frac{1}{g} \left(1-f(x)\right) \epsilon_{a\mu\rho} \frac{x^\rho}{x^2} +\frac{1}{g} A(x) \frac{x_a x_\mu}{x^2}\;,\qquad \varphi^a = \frac{1}{g} \varphi(x) \frac{x^a}{x}\,;
\ee
a possible contribution proportional to $\delta_{a\mu}$ has been set to zero by a gauge transformation and the factors of $g$ are included to simplify the following formulas. At short distances the CS deformation is unimportant and these solutions are required to approach the YM instantons. 
Given this ansatz, the euclidean bosonic action takes the form
\begin{equation}
\mc{S}_b=\mc{S}_{YM}+\mc{S}_{CS}+\mc{S}_{scalar},
\end{equation}
where
\bea
\mc{S}_{YM}&=&\dfrac{4\pi}{g^2}\int_0^\infty dx\,\left((f')^2+\dfrac{(1-f^2)^2}{2x^2}+A^2f^2\right) \nonumber\\
\mc{S}_{CS}&=&-\dfrac{4\pi\, i}{g^2}m_k\,\int_0^\infty dx\, A(1-f^2) \\
\mc{S}_{scalar}&=&\dfrac{4\pi}{g^2}\int_0^\infty dx\,\left(\ff^2f^2+\dfrac{r^2}{2}(\ff')^2\right)\,, \nonumber
\eea
and a prime denotes a derivative with respect to the euclidean distance $x$.

Integrating out $A$ sets
\beq\label{eq:InsAEoM}
A=i\dfrac{m_k(1-f^2)}{2f^2}\,,
\eneq
and the euclidean Lagrangian for the instanton becomes
\beq\label{eq:InsLag}
\mc L=\dfrac{4\pi}{g^2}\left((f')^2+\dfrac{(1-f^2)^2}{4}\Big(\dfrac{2}{x^2}+\dfrac{m_k^2}{f^2}\Big)+\ff^2 f^2+\dfrac{x^2}{2}(\ff')^2\right)\,. 
\eneq
We thus arrive at the equations of motion for the classical configuration,
\begin{subequations}\label{eq:InsEoM}
\begin{align}
f''=&\dfrac{f^2-1}{4}\left(\dfrac{4f}{x^2}+\dfrac{m_k^2}{f^3}(f^2+1)\right)+\ff^2f\,\label{eq:InsEomphi}\\
\ff''=&\dfrac{2}{x}\Big(\dfrac{\ff\,f^2}{x}-\ff'\Big)\,.\label{eq:InsEomrho}
\end{align}
\end{subequations}
As a check, at small distances the effect of the CS term is found to be unimportant and (\ref{eq:InsEoM}) admits the real solutions
\beq\label{eq:DKMTV_fs}
f(x)=\dfrac{m_g x}{\text{sinh}(m_g x)}\;,\qquad
\ff(x)=-\dfrac{1}{x}+m_g \text{coth}(m_g x)\,,
\eneq
which reproduce the instantons of the YM theory.
At long distances we require the scalar field to approach the Coulomb branch expectation value, namely $\varphi(\infty)= m_g$. In this regime, the equation of motion for $f$ is dominated by the CS term and the field approaches a limiting value
\be
f(\infty)^4= \frac{m_k^2}{m_k^2+ 4 m_g^2}\,.
\ee

We have solved this system of equations numerically for solutions that interpolate between the previous asymptotic behaviors. For nonzero CS mass, both $f(x)$ and $\varphi(x)$ are found to be slowly varying (as opposed to the highly localized instantons of the YM theory), and a numerical evaluation of the euclidean action shows that it diverges with the size of the euclidean space. The reason for this is in fact the same as in~\cite{Pisarski:1986gr}: at large $x$ the Lagrangian evaluates to
\beq\label{eq:InsLagLim}
\mathcal{L}\approx\dfrac{4\pi}{g^2}\left(\dfrac{(1-f(\infty)^2)^2}{4}\dfrac{m_k^2}{f(\infty)^2}+m_g^2 f(\infty)^2 +...\right),
\eneq
which is strictly positive, explaining the divergence of the action. 

We conclude that the instantons of the YM theory, when deformed by a CS interaction, have a divergent action. We need to look somewhere else for a mechanism to resolve singularities.
This ends our analysis of the classical YM--CS theory. In what follows we first review the appearence of Coulomb branch singularity in the UV limit of the gauge theory and then argue that the singularity is resolved by perturbative effects that become important at long distances.

\section{The UV limit: Yang--Mills theory}\label{sec:UVYM}

In this section we will calculate the one loop correction to the two-point function of $\phi$ along the Coulomb branch (\ref{eq:vev}) in the limit $m_g^2 \gg m_k^2$. This is the UV region, where the CS deformation may be ignored. In this limit the theory reduces to the 3d $SU(2)$ $\mc N=4$ theory. The quantum-corrected Coulomb branch metric was obtained in~\cite{Seiberg:1996nz}, and the  one loop calculation was carried out in~\cite{Dorey:1997ij}. However, these methods rely heavily on the $\mc N=4$ supersymmetry and it is not clear how to generalize them to include a CS deformation. For this reason, here we will reproduce the known result using an approach that can be extended to the $\mc N=2$ YM--CS theory. This will serve mostly as a warm up for our real motivation --the YM--CS theory.

\subsection{One loop calculation}\label{subsec:YM-oneloop}

Our goal is to compute the one-loop corrected two-point function
\be\label{eq:Sigmadef}
\langle \chi_3(p) \chi_3(-p) \rangle = G^{(0)}(p)+ G^{(0)}(p) \Sigma^{(1)}(p) G^{(0)}(p) + \ldots = \frac{1}{(G^{(0)})^{-1}(p)-\Sigma^{(1)}(p)}\,,
\ee
for now in the limit $m_k \to 0$. Here $G^{(0)}(p)= i/p^2$ is the tree-level propagator for $\chi_3$.
Since supersymmetry forbids the generation of a mass term, $\Sigma(p) \propto p^2$ plus higher order terms in $p^2$.  Using the fact that at one loop the self-energy is proportional to the interaction $g^2$ we have, on dimensional grounds,
\be\label{eq:Sigmadef2}
\Sigma^{(1)}(p) = i \Sigma_1 \,\frac{g^2}{m_g} \,p^2 + \mc O(p^4) =i \Sigma_1\,\frac{g}{\sqrt{2} |v|} \,p^2+ \mc O(p^4)\,.
\ee
$\Sigma_1$ is a dimensionless coefficient that will be calculated explicitly.\footnote{The index denotes the loop order, and the factor of $i$ anticipates that $\Sigma_1$ is real.} Plugging this form into (\ref{eq:Sigmadef}) and Fourier transforming back to position space gives a one-loop kinetic term
\be\label{eq:C-metric1}
L \supset \left(1 + \Sigma_1\,\frac{g}{\sqrt{2} |v|}\right) \frac{1}{2} \partial_\mu \chi_3 \partial^\mu \chi_3\,.
\ee
This defines the quantum-corrected Coulomb branch metric at the position $v$, with the rest of the components fixed by supersymmetry.

Our approach is very straightforward: we will calculate all the one loop diagrams that contribute to (\ref{eq:Sigmadef}) using the component fields and interactions described in \S \ref{subsec:classicCoulomb}. In this approach, supersymmetry is not used explicitly, although its consequences (such as the cancellation of divergences) will be seen directly. It would be nice to apply a method that takes advantage of supersymmetry from the start, perhaps in terms of the superspace background field approach of~\cite{Buchbinder:2010em,Buchbinder:2010ez,Buchbinder:2012pr}. Furthermore, our final result for the quantum corrected metric in the YM--CS theory will be so simple that it suggests that a more direct way of calculating it could exist.

We work in Landau gauge
\be
\partial^\mu A_\mu=0\,,
\ee
which is very convenient for calculations in the Coulomb phase. It can be obtained by taking the limit $\xi \to 0$ of the $R_\xi$ gauge fixing described in the Appendix. In this case there are no ghosts, the derivative couplings in (\ref{eq:Lquad}) and (\ref{eq:Lint1}) vanish, and $\chi_\alpha$ is massless. The bosonic propagators simplify to
\bea\label{eq:scalarpropsYM}
\langle A_\mu^\alpha(p) A_\nu^\beta(-p) \rangle &=& - i \delta^{\alpha \beta} \frac{g_{\mu\nu}-p_\mu p_\nu/ p^2}{p^2-m_g^2} \nonumber\\
\langle \sigma_\alpha(p) \sigma_\beta(-p) \rangle &=& i \delta_{\alpha \beta} \frac{1}{p^2-m_g^2} \nonumber\\
\langle \chi_\alpha(p) \chi_\beta(-p) \rangle &=& i \delta_{\alpha \beta} \frac{1}{p^2}\\
\langle \t \chi_\alpha(p) \t \chi_\beta(-p) \rangle &=& i \delta_{\alpha \beta} \frac{1}{p^2-m_g^2} \nonumber\,,
\eea
and for the fermions
\bea\label{eq:fermionpropsYM}
\langle \lambda_\alpha(p) \bar \lambda_\beta(p) \rangle &=& i \delta_{\alpha \beta} \frac{ \not \! p }{p^2-m_g^2} \nonumber\\
\langle \psi_\alpha(p) \bar \psi_\beta(p) \rangle &=& i \delta_{\alpha \beta} \frac{\not \! p }{p^2-m_g^2}\\
\langle \lambda_\alpha(p) \bar \psi_\beta(p) \rangle &=& i \epsilon_{\alpha \beta} m_g \frac{1}{p^2-m_g^2 } \,. \nonumber
\eea

The one loop diagrams that contribute to the $\chi_3$ two point function are as follows. The total bosonic contribution involves 7 one loop diagrams, which can be written in terms of the previous propagators to give
\bea\label{eq:YM-Sb}
\Sigma_b &= &-g^2 \int \frac{d^3 q}{(2\pi)^3} \Big[-i \langle A_\mu^\alpha A^\mu_\alpha \rangle_q +i \langle \sigma_\alpha \sigma_\alpha \rangle_q +i \langle \t \chi_\alpha \t \chi_\alpha \rangle_q +2 m_g^2 \langle A_\mu^\alpha A_\nu^\beta \rangle_q \langle A_\alpha^\mu A_\beta^\nu \rangle_{p-q} \\
&+&4  p_\mu p_\nu \epsilon_{\alpha \beta} \epsilon_{\gamma \delta}  \langle A^\mu_\alpha A^\nu_\gamma \rangle_q \langle \chi_\beta \chi_\delta \rangle_{p-q}+ 2m_g^2 \langle \sigma_\alpha \sigma_\beta \rangle_q \langle \sigma_\alpha \sigma_\beta\rangle_{p-q}+ 2m_g^2\langle \t \chi_\alpha \t \chi_\beta \rangle_q \langle \t \chi_\alpha \t \chi_\beta\rangle_{p-q}\Big]\nonumber\,.
\eea
We use the shorthand notation $\langle \phi_1 \phi_2 \rangle_q \equiv \langle \phi_1(q) \phi_2(-q) \rangle $. The fermionic contribution amounts to
\be\label{eq:YM-Sf}
\Sigma_f = 2g^2 \epsilon_{\alpha \beta}\epsilon_{\gamma \delta} \int \frac{d^3 q}{(2\pi)^3}\Big[ \langle \lambda_\alpha \bar \lambda_\gamma\rangle_q  \langle \psi_\beta \bar \psi_\delta\rangle_{q-p} + \langle \lambda_\alpha\bar \psi_\delta \rangle_q   \langle \lambda_\gamma \bar \psi_\beta \rangle_{q-p} \Big] \,.
\ee
Individual diagrams have linear divergences, but they exactly cancel in the total contribution $\Sigma = \Sigma_b + \Sigma_f$, so no regulator is needed at this order. This is expected from supersymmetry. Expanding the self-energy in powers of $p^2$ and performing the loop integrals obtains, 
\be\label{eq:Sigma1YM}
\Sigma^{(1)} = - \frac{i}{2\pi} \frac{g^2}{m_g} p^2 + \mc O (p^4)\,.
\ee

We conclude that $\Sigma_1 = -1/(2\pi)$ --- see (\ref{eq:Sigmadef2}) --- and the quantum Coulomb branch metric is (restoring the real and imaginary parts of the flat direction)
\be
L \supset \left(1 - \frac{1}{2\sqrt{2} \pi}\,\frac{g}{ |\phi_3|}\right) \partial_\mu \phi_3 \partial^\mu \bar \phi_3\,,
\ee
which reproduces~\cite{Seiberg:1996nz,Dorey:1997ij}. This result is exact in perturbation theory, and signals a perturbative singularity at
$$
|\phi_3|=\frac{1}{2\sqrt{2} \pi} g\,.
$$

\section{Quantum effects in pure Chern--Simons theory}\label{sec:CS}

Having understood the perturbative singularity in the YM limit, we will next focus on the deep IR. In this limit, the CS deformation dominates over the kinetic terms for the vector superfield. So in this section we will determine the quantum corrected Coulomb branch metric of the Chern--Simons-matter theory which is also interesting in its own right.\footnote{We refer the reader to~\cite{Dunne:1998qy} for a detailed review of Chern--Simons gauge theories with references to the original literature. Some of the early works on supersymmetric CS theories are~\cite{Hlousek:1990jf,Ivanov:1991fn,Avdeev:1992jt,Kao:1995gf}; also see~\cite{Gaiotto:2007qi} for a more recent review.}

The resulting theory is obtained from (\ref{eq:L1}) by rescaling
$$
(A_\mu, \sigma, \lambda, D) \to \frac{1}{g} (A_\mu, \sigma, \lambda, D)
$$
and then taking the limit $g \to \infty$. Now $\sigma$ and $\lambda$ also become auxiliary, and integrating them out obtains
\bea\label{eq:LpureCS}
L& =& (D_\mu \phi_i)^\dag D^\mu \phi_i+ i \bar \psi_i \not  \!  \! D  \psi_i+\frac{k}{8 \pi} \,\epsilon^{\mu \nu \rho} (A_\mu^a \partial_\nu A_\rho^a +\frac{1}{3} f^{abc} A_\mu^a A_\nu^b A_\rho^c)  \\
&-& \frac{16\pi^2}{k^2} (\phi^\dag T^a T^b \phi) (\phi^\dag T^a \phi) (\phi^\dag T^b \phi)+ \frac{4\pi}{k} (\bar \psi T^a \psi) (\phi^\dag T^a \phi) + \frac{8\pi}{k} (\bar \psi T^a \phi) (\phi^\dag T^a \psi)\nonumber\,.
\eea
More details are given in \S \ref{subsec:3dsusy}. It is sometimes convenient to introduce the parameter
\be\label{eq:defkappa}
\kappa \equiv \frac{k}{4\pi}\,.
\ee
Perturbation theory is an expansion in powers of $\kappa^{-1}$. 

This theory has a one-dimensional Coulomb branch parametrized by $\tr (\phi^2)$. The kinetic term is not protected against quantum corrections and, unlike the case of $\mc N=4$ supersymmetry, it receives corrections to all orders in perturbation theory. Here we will calculate the one loop corrections and show that there is no singularity. This finite effect is the dominant contribution in the perturbative regime $k \gg 4\pi$, which we henceforth assume.\footnote{There are also interesting quantum corrections to the CS level of the $U(1)$ gauge field; see e.g.~\cite{Spiridonov:1990tx,Chen:1994zx}.}\\

Let us develop some intuition about the corrections that can appear. The crucial difference between the CS matter theory (\ref{eq:LpureCS}) and the previous case with nonzero gauge coupling is that, at least in perturbation theory, there is no dimensionful parameter. The CS level $k$, which is dimensionless and quantized,
determines all the interactions. Next, turning on a Coulomb branch expectation value $v$, this will be the only dimensionful parameter, and hence it cannot appear in the loop corrected kinetic term for $\phi_3$. In the Yang--Mills case there was a dimensionless parameter $v/g$ that was appearing in loop corrections, but this is no longer possible in the pure CS-matter case. We thus conclude that, if there are quantum corrections to the Coulomb branch metric, they have to be independent of $v$. This argument explains the absence of singularities along the Coulomb branch at the perturbative level.

\subsection{One loop quantum corrections}\label{subsec:CS-oneloop}

Now we need to determine whether such corrections occur and what their numerical value is.  We turn on an expectation value (\ref{eq:vev}) and compute the two-point function for the massless field $\chi_3$ --- recall our definition (\ref{eq:chidef}). At one loop, $\chi_3$ interacts with the fields $(A_\mu^\alpha, \chi_\alpha, \t \chi_\alpha, \psi_\alpha)$, where $\alpha=1,2$ are the color directions perpendicular to the Coulomb branch expectation value.

In Landau gauge $\partial_\mu A^\mu=0$, the quadratic Lagrangian for the fields $(A_\mu^\alpha, \chi_\alpha, \t \chi_\alpha, \psi_\alpha)$ is
\be\label{eq:L2CS}
L_{quad}= \frac{\kappa}{2} A_\mu^\alpha \left( \epsilon^{\mu\nu\rho} \partial_\nu + m_H g^{\mu\rho} \right) A_\rho^\alpha- \frac{1}{2} \t \chi_\alpha (\Box + m_H^2) \t \chi_\alpha- \frac{1}{2} \chi_\alpha \Box  \chi_\alpha +\bar \psi_\alpha (i\not  \! \partial+m_H)  \psi_\alpha\,,
\ee
where the mass induced by the Higgs mechanism is
\be
m_H \equiv \frac{2 v^2}{\kappa}\,.
\ee
From this quadratic action we can read off the propagators
\bea
\langle A_\mu^\alpha(p) A_\nu^\beta(-p) \rangle &=& - i \,\delta^{\alpha \beta}\,\kappa^{-1}\, \frac{m_H (g_{\mu\nu} - p_\mu p_\nu/p^2) - i \epsilon_{\mu\nu\rho} p^\rho}{p^2 - m_H^2}\nonumber\\
\langle \chi_\alpha(p) \chi_\beta(-p) \rangle &=& \delta_{\alpha \beta}\frac{i}{p^2}\;,\qquad\langle \t\chi_\alpha(p) \t\chi_\beta(-p) \rangle = \delta_{\alpha \beta}\frac{i}{p^2 - m_H^2} \nonumber\\
\langle \psi_\alpha(p) \bar \psi_\beta(p) \rangle &=&\delta_{\alpha \beta}\frac{i(\not \! p+ m_H)}{p^2 - m_H^2} \,.
\eea
Furthermore, the interaction terms that contribute to the one-loop action of $\phi_3$ are
\bea
L_{int}&=&- \frac{1}{\kappa^2} \left(2 \sqrt{2}v^3 \t\chi_\alpha(- \chi_\alpha \t \chi_3+ 2 \t\chi_\alpha \chi_3) + 6 v^2 \t \chi_\alpha \chi_3 (-\chi_\alpha \t \chi_3 + \t \chi_\alpha \chi_3 )+ v^2(\chi_\alpha^2 + \t \chi_\alpha^2)\t \chi_3^2 \right)\nonumber \\
&+& \left(\frac{2}{\kappa} \bar \psi_\alpha \psi_\alpha+ (A_\mu^\alpha)^2\right)
\left( \sqrt{2} v \chi_3 +\frac{1}{2} \chi_3^2 + \frac{1}{2}\t \chi_3^2\right) +2 \epsilon_{\alpha \beta} A^\mu_\alpha (\chi_\beta \partial_\mu \chi_3+ \t \chi_\beta \partial_\mu \t \chi_3) \,.
\eea

Now we are ready to compute the one loop self-energy for $\chi_3$. The bosonic contributions sum up to
\bea\label{eq:CS-Sb}
\Sigma_b &=&- \int \frac{d^3 q}{(2\pi)^3} \Big(i \frac{6}{\kappa}\,m_H  \langle \t \chi_\alpha \t \chi_\alpha \rangle_q-i\langle A_\mu^\alpha A^\mu_\alpha \rangle_q + \frac{8}{\kappa} \,m_H^3 \langle \t \chi_\alpha \t \chi_\beta \rangle_q \langle \t \chi_\alpha \t \chi_\beta \rangle_{p-q}\nonumber\\
&&\qquad \qquad +4 p_\mu p_\nu \epsilon_{\alpha \beta} \epsilon_{\gamma \delta}\langle A^\mu_\alpha A^\nu_\gamma \rangle_q \langle \chi_\beta \chi_\delta \rangle_{p-q} +4 v^2\langle A_\mu^\alpha A_\nu^\beta \rangle_q \langle A_\alpha^\mu A_\beta^\nu \rangle_{p-q}\Big)\,.
\eea
This is of the same form as (\ref{eq:YM-Sb}) (after a rescaling to absorb the powers of $g$), except that $\sigma$ does not appear any more because it is an auxiliary field in the CS theory. On the other hand, the fermionic contributions are
\be\label{eq:CS-Sf}
\Sigma_f = \frac{2}{\kappa} \, \int \frac{d^3 q}{(2\pi)^3} \left(i  \langle \psi_\alpha \bar \psi_\alpha\rangle_q -2 m_H  \langle \psi_\alpha \bar \psi_\beta\rangle_q  \langle \psi_\alpha \bar \psi_\beta\rangle_{p-q}\right)\,.
\ee

Replacing the previous expressions for the propagators and expanding in powers of $p^2$ obtains
\be\label{eq:CSmres}
\Sigma^{(1)}= \Sigma_b+\Sigma_f = - \frac{i}{2\pi \kappa} p^2 + \mc O(p^4)\,.
\ee
Thus, the Coulomb branch metric including one loop effects becomes
\be
L \supset \left(1- \frac{1}{2\pi \kappa} \right) \partial_\mu \phi_3 \partial^\mu \bar \phi_3\,.
\ee
In summary, in the CS-matter theory there is no Coulomb branch singularity, and instead there is a finite one loop correction to the metric proportional to $\kappa^{-1}$ and independent of $\phi_3$. We also expect nonzero higher loop corrections, suppressed by higher powers of the coupling, so our result gives a good approximation in the perturbative regime $\kappa^{-1} \ll 2\pi$.

\section{The quantum Coulomb branch of YM--CS theory}\label{sec:fullYMCS}

Finally, we are ready to attack the full problem of computing the one-loop corrected Coulomb branch metric for the YM--CS theory (\ref{eq:L1}). The result hasn't been obtained before, and the calculations are somewhat involved, so it is important to perform consistency checks. Therefore, instead of specializing to the Landau gauge, here we will work in an arbitrary $R_\xi$ gauge and will show that the self-energy is independent of the gauge fixing parameter. This provides a nontrivial verification on our result, implying its gauge invariance and the restoration of supersymmetry (which is broken by the gauge fixing function). Other consistency checks will include the exact cancellation of divergences and the correct UV behavior.

\subsection{One loop metric along the Coulomb branch}\label{subsec:oneloopYMCS}

The $R_\xi$ gauge corresponds to the gauge-fixing function (see e.g.~\cite{Peskin:1995ev})
\be
G^a = \partial_\mu A^{a\,\mu} - i g \xi \left(\langle \phi^\dag_i \rangle T^a_{ij} \delta \phi_j - \delta \phi^\dag_i T^a_{ij} \langle \phi_j \rangle \right)\;,\qquad L_{g.f.}= - \frac{1}{2\xi} (G^a)^2
\ee
where $\delta \phi$ is the fluctuation around the vacuum expectation value. Evaluating this and the Faddeev--Popov determinant for our theory, we find that the gauge fixing and ghost Lagrangian terms that contribute at one loop are
\be
L_{g.f.} +L_{ghost}\supset - \frac{1}{2 \xi} (\partial_\mu A^\mu_\alpha)^2 - m_g \epsilon_{\alpha \beta} \partial_\mu A^\mu_\alpha \chi_\beta - \frac{1}{2} \xi m_g^2 \chi_\alpha^2 - \bar c^\alpha (\Box + \xi m_g^2 + \xi g m_g \chi_3) c^\alpha\,.
\ee
More details may be found in \S \ref{subsec:app-propagators}.
The fields with color index parallel to the color-breaking direction $a=3$ do not contribute to the one loop effective action. Recall that $m_g = \sqrt{2} g v$, and that we are working in the convention where $g$ appears in the interactions and not in front of $F_{\mu\nu}^2$. 

The propagators for the massive fields for arbitrary $\xi$ are given in the Appendix.
Also, note that in the $R_\xi$ gauge the derivative interaction term $L \supset g \epsilon_{\alpha \beta} \partial_\mu A^\mu_\alpha \chi_\beta \chi_3$ also contributes to the one loop self-energy.

There are 13 one loop bosonic contributions to the $\chi_3$ two-point function,
\bea
\Sigma_b &=& - g^2 \int \frac{d^3 q}{(2\pi)^3} \Big\{-i \langle A_\mu^\alpha A^\mu_\alpha \rangle_q +i \langle \sigma_\alpha \sigma_\alpha \rangle_q +i \langle \t \chi_\alpha \t \chi_\alpha \rangle_q  +2 m_g^2 \langle A_\mu^\alpha A_\nu^\beta \rangle_q \langle A_\alpha^\mu A_\beta^\nu \rangle_{p-q} \nonumber\\
&+&2m_g^2 \Big[\langle \sigma_\alpha \sigma_\beta \rangle_q \langle \sigma_\alpha \sigma_\beta\rangle_{p-q}+\langle \t \chi_\alpha \t \chi_\beta \rangle_q \langle \t \chi_\alpha \t \chi_\beta\rangle_{p-q}+ 2\langle \sigma_\alpha \t \chi_\beta \rangle_q \langle \sigma_\alpha \t \chi_\beta \rangle_{p-q} \Big]\nonumber\\
&+&m_k^2\,\epsilon_{\alpha \beta}\epsilon_{\gamma \delta}\Big[ \langle \sigma_\alpha \sigma_\gamma\rangle_q  \langle \t \chi_\beta \t \chi_\delta\rangle_{p-q} + \langle \sigma_\alpha\t \chi_\delta\rangle_q  \langle \sigma_\gamma \t \chi_\beta\rangle_{p-q}  \Big]-4 m_g m_k \epsilon_{\alpha \beta}
\langle \sigma_\alpha \sigma_\gamma \rangle_q \langle \sigma_\gamma \t \chi_\beta \rangle_{p-q} \nonumber\\
&-&4 m_g m_k \epsilon_{\alpha \beta}\langle \t \chi_\beta \t \chi_\gamma \rangle_q \langle \sigma_\alpha \t \chi_\gamma \rangle_{p-q}+\epsilon_{\alpha \beta} \epsilon_{\gamma \delta}\,(4 p_\mu p_\nu+q_\mu q_\nu) \langle A^\mu_\alpha A^\nu_\gamma \rangle_q \langle \chi_\beta \chi_\delta \rangle_{p-q} \Big\}\,.
\eea
The fermionic contributions are simpler,
\be
\Sigma_f = 2g^2 \epsilon_{\alpha \beta}\epsilon_{\gamma \delta} \int_q \Big[ \langle \lambda_\alpha \bar \lambda_\gamma\rangle_q  \langle \psi_\beta \bar \psi_\delta\rangle_{q-p} + \langle \lambda_\alpha\bar \psi_\delta \rangle_q   \langle \lambda_\gamma \bar \psi_\beta \rangle_{q-p} \Big]\,.
\ee
We do not write explicitly the propagators (that can be found in the Appendix) in order to exhibit explicitly the various symmetry factors and couplings. The ghost contribution can be obtained from Feynman diagrams or directly from the Faddeev--Popov determinant $\det (\delta_\alpha G_\beta) \propto \det{}^2 (\Box + \xi m_g^2 + g \xi m_g \chi_3)$, with the result
\be
\Sigma_{ghost}= -2 g^2 \xi^2 m_g^2 \int_q \frac{1}{q^2 - \xi m_g^2}\frac{1}{(p-q)^2 - \xi m_g^2}\,.
\ee

Putting these results together, we finally arrive at the total one loop self-energy
\be\label{eq:YMCSmres}
\Sigma^{(1)} = \Sigma_b + \Sigma_f + \Sigma_{ghost} = - \frac{i}{\pi} \frac{g^2}{m_+ + m_-} p^2 + \mc O(p^4)\,,
\ee
where the mass eigenvalues $m_{\pm}^2$ were defined in (\ref{eq:mpm}).
As promised, this is independent of the gauge fixing parameter $\xi$, thus verifying the gauge invariance of the self-energy. This is a strikingly simple result: it is the same as the pure YM quantum correction (\ref{eq:Sigma1YM}) after replacing $m_g$ by the averaged mass  $(m_+ + m_-)/2$ of the massive modes! 
Of course, an important difference to keep in mind is that the Coulomb branch metric of YM--CS is expected to receive higher loop corrections, while the YM result is exact in perturbation theory.

\begin{figure}[t!]
\begin{center}
\includegraphics[width=0.6\textwidth]{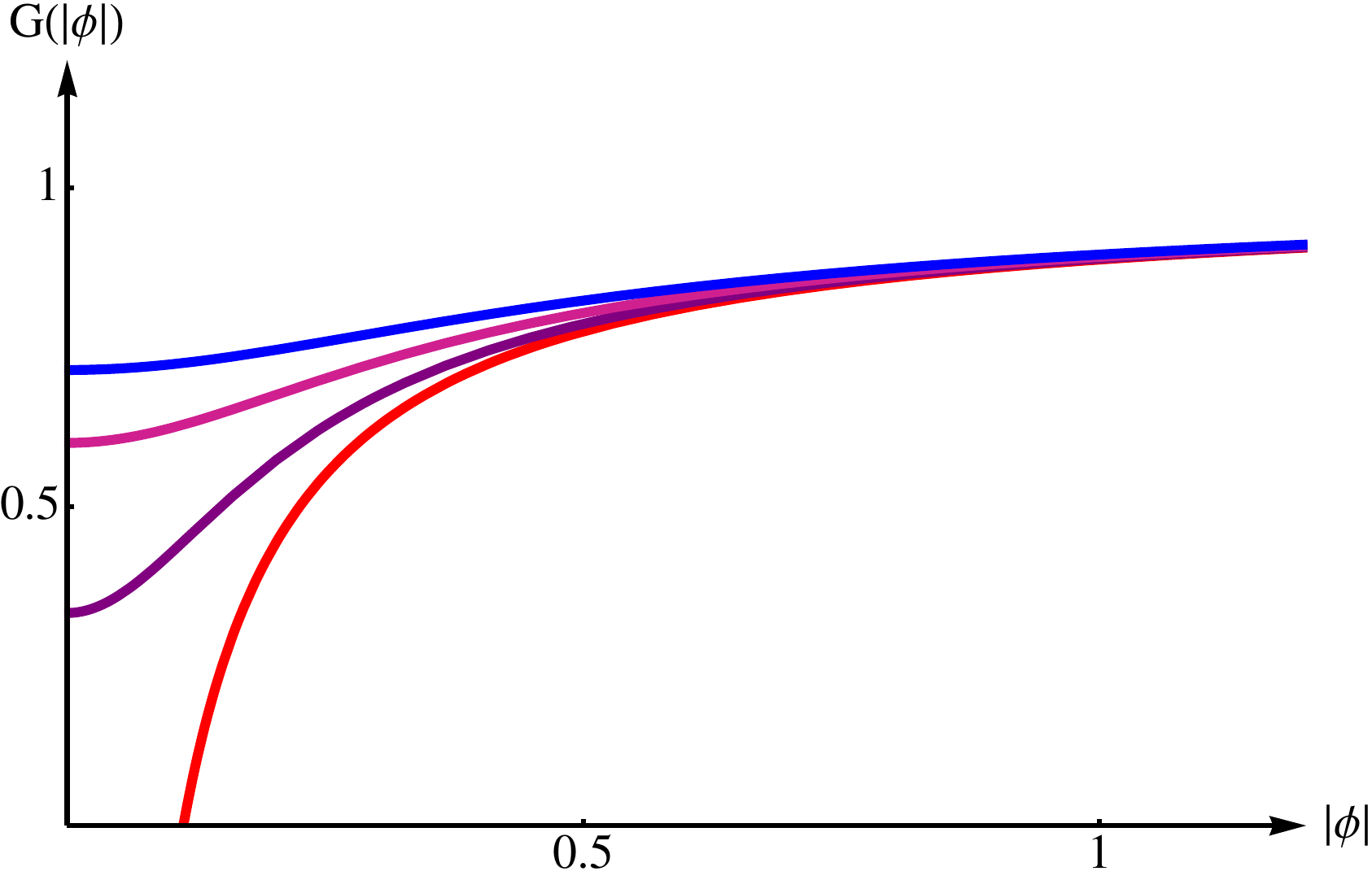}
\caption{Plot of the function $G(|\phi|)$ in the one loop Coulomb branch metric in YM--CS (namely, $S\supset \int G(|\phi|) \partial_\mu \phi_3 \partial^\mu \bar \phi_3$), for different values of $k$. The four curves correspond to $k=0,6,10,14$ (in red, violet, magenta and blue, respectively), with $g=1$ in all cases. For $k=0$, the metric vanishes at $\phi= \frac{g}{2 \sqrt{2}\pi}$, while it stays  regular everywhere along the Coulomb branch for $k$ large enough that our perturbative computations can be trusted.}
\label{fig:plot1}
\end{center}
\end{figure}

In summary, the Coulomb branch metric including one loop corrections is
\be 
S \supset \int d^3 x \left(1- \frac{1}{2\pi} \frac{1}{\sqrt{2 \frac{|\phi_3|^2}{g^2}+ \frac{\kappa^2}{4}}} \right) \partial_\mu \phi_3 \partial^\mu \bar \phi_3\,.
\ee
This is our main result.
In the UV regime $|\phi| \gg \kappa g$ this recovers the known YM result, providing another test for our result. When $|\phi| \sim \kappa g$ the effects from the CS deformation become important and for $|\phi| \ll \kappa g$ the metric flows to a constant value $G = 1- \frac{1}{\pi \kappa}$. When $\kappa=\frac k{4\pi}\gg 1$ (which is when our perturbative computations can be trusted), $G>0$. Therefore, the perturbative singularity of the YM theory is resolved by the topological mass. The full metric $G(|\phi|)$ interpolating between the UV and IR for different values of $\kappa$ is shown in Figure \ref{fig:plot1}, together with the one loop result in the pure YM case.

\subsection{Nondecoupling of massive modes}\label{subsec:nondecoupling}

Finally, let us compare the IR limit $g \to \infty$ of the full YM--CS calculation with the result in the CS theory of \S \ref{sec:CS}. When the gauge coupling is taken to infinity,
\beq\label{eq:mpmginty}
m_+\,\to\, m_k\sim g^2\;,\qquad m_- \,\to\, m_H\sim O(g^0)\,,
\eneq
and hence \reff{eq:YMCSmres} and \reff{eq:CSmres} are related by
\beq
\begin{split}
\Sigma^{(1)}_{\text{YM+CS+m}}
\IR&-\dfrac{i\,g^2}{\pi\,m_k }p^2 + \mc O(p^4)\\
=&2\,\Sigma^{(1)}_{\text{CS+m}}+ \mc O(p^4)\,.
\end{split}
\eneq
Therefore, the YM--CS and CS results differ by a factor of 2. At first this is a bit puzzling, because the CS interaction dominates over the YM kinetic term in the IR; so we would expect the CS calculation to reproduce the long distance behavior of the full theory.

In order to understand this difference consider, for instance, the one loop contribution from the interaction term involving $\chi_\alpha$ and $A^\mu_\beta$. In Landau gauge, 
\beq\label{eq:SigmaChiAYM}
\Sigma_{YM+CS+m}^{(1)}[\chi A]= -i\dfrac{16g^2}{3}p^2\int\dfrac{dq^3}{(2\pi)^3}\dfrac{(q^2-m_g^2)}{(p-q)^2(q^2-m_+^2)(q^2-m_-^2)}\,,
\eneq
while for the CS+matter theory,
\beq\label{eq:SigmaChiA}
\Sigma_{CS+m}^{(1)}[\chi A]= -i\dfrac{16}{3\kappa}p^2\int\dfrac{dq^3}{(2\pi)^3}\dfrac{m_H}{(p-q)^2(q^2-m_H^2)}\,.
\eneq
Taking the limit $g\rightarrow\infty$ \emph{before} performing the integral, both contributions give the same result. However, performing the integrals at finite $g$ and taking the IR limit at the end of the calculation gives a different answer. Let us see how this comes about.

Wick-rotating and expanding to order $p^2$ obtains
\beq\label{eq:decouple0}
\Sigma_{YM+CS+m}^{(1)}[\chi A]=-i\dfrac{4g^2}{3\pi^2}p^2\int_{-\infty}^{+\infty}dq \dfrac{(q^2+m_g^2)}{(q^2+m_+^2)(q^2+m_-^2)} + \mc O(p^4)\,.
\eneq
This integral can be done using residue methods. For instance, closing the contour in the upper complex plane, the two poles that contribute are at $q=i\,m_\pm$, obtaining
\be\label{eq:decouple1}
\Sigma_{YM+CS+m}^{(1)}[\chi A]=-\dfrac{8g^2}{3\pi}i\dfrac{p^2}{m_++m_-}+ \mc O(p^4)\IR-\dfrac{8i\, p^2}{3\pi\kappa}+ \mc O(p^4)\,.
\ee
In contrast, in the CS+matter theory there is only one pole that contributes:
\beq\label{eq:decouple2}
\Sigma_{CS+m}^{(1)}[\chi A]=-i\dfrac{4}{3\kappa\,\pi^2}p^2\int_{-\infty}^{+\infty}dq \dfrac{m_H}{q^2+m_H^2} + \mc O(p^4)=-\dfrac{4i\,p^2}{3\pi \kappa}+ \mc O(p^4)\,.
\eneq
Thus, the relative factor of $2$ comes from the massive states that yield the extra pole at $q^2 = -m_+^2$, which is absent in the CS theory. These masses diverge for $g \to \infty$ but, as we just found, they give a finite wavefunction renormalization to the Coulomb branch field. If we instead take $g \to \infty$ first and then perform the integral both answers agree because the massive pole goes to infinity and does not contribute to the contour integral.

It is interesting that the Coulomb branch metric is in this way sensitive to the UV completion. This is related to anomalies. In three dimensions, the generation of a CS interaction from massive fermions is itself an example of this~\cite{Redlich:1983kn,Niemi:1983rq,Redlich:1983dv}, related to the parity anomaly. The nondecoupling finite effect on the Coulomb branch that we just described is associated to the trace anomaly.\footnote{An analog of this in four dimensions that is phenomenologically important is the triangle diagram for Higgs production from gluon fusion or the $h \to \gamma \gamma$ decay, where loops of very massive fermions give finite contributions~\cite{Donoghue:1992dd}. We thank M. Peskin for discussions on these points.}

\section*{Acknowledgments}
We are grateful to 
M.S.~Bianchi,
G.~Dunne,
S.~Kachru,
C.~Oleari,
M.~Peskin, 
S.~Penati,
D.~Rosa,
M.~Siani
for helpful
discussions.
A.T.~and F.S.~have been supported in part by INFN, by the MIUR-FIRB grant RBFR10QS5J ``String Theory and Fundamental Interactions'', and by the MIUR-PRIN contract 2009-KHZKRX. The research of A.T. is also supported by the ERC Starting Grant 307286 (XD-STRING). The research of G.T.~is supported in part by the National Science Foundation under grant no.~PHY-0756174. 

\appendix

\section{Some useful field theory results}\label{app:conventions}

In this Appendix we collect our field theory conventions and other required results for the calculations in the main part of the paper.

The metric signature is $(+--)$; 3d fermions are Dirac, with $\bar \psi = \psi^\dag \gamma^0$. A possible representation of gamma matrices is $\gamma^0=\sigma_2,\,\gamma^1= i \sigma_3,\,\gamma^2 = i \sigma_1$. The covariant derivative is given by
\be
D_\mu \phi_i = \partial_\mu \phi_i - i A_\mu^a T^a_{ij} \phi_j\,.
\ee
For the adjoint representation, $T^a_{ij} = - i f^{aij}$ and $D_\mu \phi_a = \partial_\mu \phi_a + f^{abc} A^b_\mu \phi_c$.

\subsection{Three-dimensional supersymmetric theories}\label{subsec:3dsusy}

In three dimensions, the vector superfield contains the gauge field $A_\mu^a$ and gaugino $\lambda^a$, a real scalar $\sigma^a$ (the extra component of the gauge field in reducing from 4d to 3d) and an auxiliary D-term $D^a$. Their lagrangian is
\be\label{eq:Lgauge}
L_{gauge}= \frac{1}{g^2} \left[-\frac{1}{4} F_{\mu\nu}^a F^{a\mu\nu}+ \frac{1}{2} D_\mu \sigma^a D^\mu \sigma^a + i \bar \lambda^a \not\!\! D \lambda^a -\bar \lambda \sigma \lambda+ \frac{1}{2} D^a D^a \right]\,.
\ee
The gaugino is a 3d Dirac fermion.

Moving to the matter sector, a chiral superfield contains a complex scalar $\phi$, a Dirac fermion $\psi$ and an auxiliary field $F$. The lagrangian reads
\bea\label{eq:Lmatter}
L_{matter}&=& (D_\mu \phi_i)^\dag D^\mu \phi_i + i \bar \psi_i \not \!\! D \psi_i - \phi^\dag \sigma^2 \phi + \phi^\dag D \phi -\bar \psi \sigma \psi +i \sqrt{2} \phi^\dag \bar \lambda \psi - i \sqrt{2}  \bar\psi \lambda \phi \nonumber\\
&+&F^\dag_i F_i + F_i \frac{\partial W}{\partial \phi_i} +F_i^\dag \left(\frac{\partial W}{\partial \phi_i}\right)^\dag - \frac{1}{2} \frac{\partial^2 W}{\partial \phi_i \partial \phi_j}\psi_i \psi_j - \frac{1}{2}\left(\frac{\partial^2 W}{\partial \phi_i \partial \phi_j}\right)^\dag \bar \psi_i \bar\psi_j\,.
\eea
The fields from the vector superfield act on the matter ones as matrices. For instance, $\phi^\dag D \phi \equiv \phi_i^\dag (T^a D^a)_{ij} \phi_j$. Similarly,
$\phi^\dag \sigma^2 \phi \equiv \phi^\dag_i (\sigma^a T^a_{ij}) (\sigma^b T^b_{jk}) \phi_k$.
Integrating out the D-term sets $D^a = -g^2 \phi^\dag T^a \phi$, and the relevant part of the lagrangian becomes
\be
\frac{1}{2g^2} D^a D^a + \phi^\dag D \phi \to - \frac{g^2}{2}  (\phi^\dag T^a \phi) (\phi^\dag T^a \phi)\,.
\ee

Consider now the 3d $\mc N=4$ theory with gauge group $SU(N)$ and no flavors. This arises for the special case of an $\mc N=2$ theory with a single matter superfield in the adjoint representation, and vanishing superpotential, $W=0$. Therefore, the Lagrangian reads
\bea
L_{\mc N=4}&=&\frac{1}{g^2} \left[-\frac{1}{4} F_{\mu\nu}^a F^{a\mu\nu}+ \frac{1}{2} D_\mu \sigma^a D^\mu \sigma^a + i \bar \lambda^a \not\!\! D \lambda^a -\bar \lambda \sigma \lambda \right]+(D_\mu \phi^a)^\dag D^\mu \phi^a+i \bar \psi^a \not\! D \psi^a \nonumber\\
&-& \bar \lambda \sigma \lambda - \bar \psi \sigma \psi - \phi^\dag \sigma^2 \phi + i \sqrt{2}\,\phi^\dag \bar \lambda \psi - i \sqrt{2}\,  \bar \psi \lambda \phi - \frac{g^2}{2} (\phi^\dag T^a \phi)(\phi^\dag T^a \phi)\,.
\eea
Here we have included a factor of $1/g^2$ in the vector multiplet kinetic terms, while in the main part of the paper these kinetic terms are taken to be canonical. The convention here is more convenient for understanding the $g \to \infty$ limit, which we will consider shortly. On the other hand, the choice of canonical kinetic terms simplifies the Feynman diagram calculations in the Coulomb branch of the full YM--CS theory.

Next, let us add an $\mc N=2$ CS deformation --- preserving more supersymmetries would lift the Coulomb branch. This is given by
\be\label{eq:Lcs}
L_{CS}= \frac{k}{8 \pi} \left(\epsilon^{\mu \nu \rho} (A_\mu^a \partial_\nu A_\rho^a + \frac{1}{3} f^{abc} A_\mu^a A_\nu^b A_\rho^c) -2 \bar \lambda^a \lambda^a + 2 D^a \sigma^a \right)\,,
\ee
so that the YM--CS theory is
\be
L = L_{\mc N=4} + L_{CS}\,.
\ee
The final form for the Lagrangian is obtained by integrating out the auxiliary D-term,
$$
D^a = - \frac{g^2 k}{4\pi} \sigma^a - g^2 \phi_i^\dag T^a_{ij} \phi_j\,.
$$
We then arrive at
\bea\label{eq:LCSfinal}
L&=&\frac{1}{g^2}\left[-\frac{1}{4} F_{\mu\nu}^a F^{a\mu\nu}+ \frac{1}{2} (D_\mu \sigma^a)^2 + i \bar \lambda^a \not\! \! D \lambda^a -\bar \lambda \sigma \lambda \right] \nonumber\\
&+&\frac{k}{8 \pi} \left[\epsilon^{\mu \nu \rho} (A_\mu^a \partial_\nu A_\rho^a + \frac{1}{3} f^{abc} A_\mu^a A_\nu^b A_\rho^c) -2 \bar \lambda^a \lambda^a  \right]-\frac{1}{2} g^2 \left( \frac{ k}{4\pi} \sigma^a +\phi_i^\dag T^a_{ij} \phi_j\right)^2 \nonumber\\
&+& (D_\mu \phi_i)^\dag (D^\mu \phi_i)+ i \bar \psi_i \not  \!  \! D  \psi_i - \phi^\dag \sigma^2 \phi - \bar \psi \sigma \psi + \sqrt{2} i (\phi^\dag \bar \lambda \psi - \bar \psi \lambda \phi)\,.
\eea
This shows directly that in the presence of $L_{CS}$, the gaugino and scalar $\sigma$ acquire a mass
\be\label{eq:mcs}
m_k \equiv \frac{g^2 k}{4\pi}\,.
\ee
Calculating the gauge field propagator requires choosing a gauge, as we will discuss momentarily. The theory studied in this paper is given by  (\ref{eq:LCSfinal}) with $SU(2)$ gauge group.

Finally, we consider the limit $g^2 \to \infty$, where the kinetic terms for the vector supermultiplet are set to zero. Now $\sigma$, $D$ and $\lambda$ are all auxiliary, and integrating them out sets
$$
\lambda^a = \frac{4\pi}{k} \sqrt{2}i (\phi^\dag T^a \psi)\;,\qquad\sigma^a = -\frac{4\pi}{k} (\phi^\dag T^a \phi)
$$
where $(\phi^\dag T^a \phi)= \phi^\dag_i T^a_{ij} \phi_j$. Therefore, the Lagrangian becomes
\bea
L& =& (D_\mu \phi_i)^\dag D^\mu \phi_i+ i \bar \psi_i \not  \!  \! D  \psi_i+\frac{k}{8 \pi} \,\epsilon^{\mu \nu \rho} (A_\mu^a \partial_\nu A_\rho^a - \frac{1}{6} f^{abc} A_\mu^a A_\nu^b A_\rho^c)  \\
&-& \frac{16\pi^2}{k^2} (\phi^\dag T^a T^b \phi) (\phi^\dag T^a \phi) (\phi^\dag T^b \phi)+ \frac{4\pi}{k} (\bar \psi T^a \psi) (\phi^\dag T^a \phi) + \frac{8\pi}{k} (\bar \psi T^a \phi) (\phi^\dag T^a \psi)\nonumber\,.
\eea

\subsection{Fields and propagators in the Coulomb phase}\label{subsec:app-propagators}

In this work we are interested in the quantum corrections to the Coulomb branch metric. The classical Coulomb branch was described in \S \ref{subsec:classicCoulomb}. It can be parametrized by the expectation value
\be
\langle \phi^a \rangle = v \delta^{a3}\;,\qquad v \in \mathbb R\,.
\ee
Let us focus on the fields that contribute to the one loop effective action for $\phi_3$, which have color indices $\alpha=1,\,2$ perpendicular to the color breaking direction $a=3$. Here we work with canonical kinetic terms for the gauge field and gauginos, which simplifies the one loop calculations.

We consider a general $R_\xi$ gauge (see~\cite{Peskin:1995ev})
\be
L_{g.f.}= - \frac{1}{2\xi} \left( \partial_\mu A^{a\,\mu} - i g \xi \left(\langle \phi^\dag_i \rangle T^a_{ij} \delta \phi_j - \delta \phi^\dag_i T^a_{ij} \langle \phi_j \rangle \right)\right)^2
\ee
where $\delta \phi$ is the fluctuation around the vacuum expectation value. In our case, the gauge fixing terms and ghost Lagrangian simplify to
\be\label{eq:Lgfghost}
L_{g.f.}+ L_{ghost}=- \frac{1}{2\xi} \left( \partial_\mu A^{a\,\mu} +m_g \xi \epsilon^{ab3} \chi_b \right)^2 - \bar c^a \left((\partial_\mu D^\mu)^{ab}+ m_g^2 \xi \delta^{ab} \right) c^b- m_g g \xi (\bar c^\alpha c^\alpha \chi_3 - \bar \chi^\alpha c^3 \chi_\alpha)\,.
\ee
The Lagrangian for the fields along along the $\alpha$ direction also involves the quadratic terms
(\ref{eq:Lquad}) and interactions (\ref{eq:Lint1}). 

The gauge fixing Lagrangian (\ref{eq:Lgfghost}) cancels the quadratic coupling between $\partial_\mu A^\mu_\alpha$ and the Goldstone mode, but the cubic interaction $\partial_\mu A^\mu_\alpha \chi_3 \chi_\beta$ survives. It is also possible to choose a quadratic gauge fixing function
$$
G^a=\partial_\mu A^{a\,\mu} - i g \xi \left( \phi^\dag_i  T^a_{ij}  \phi_j -  \phi^\dag_i T^a_{ij}  \phi_j  \right)
$$
that also cancels the cubic derivative interaction and has a simple Faddeev--Popov determinant $\det(\delta_\beta G_\alpha) \propto \det^2(\Box+ 2 g^2 \xi |\phi_3|^2)$. We have verified that both gauge fixing functions give the same result.

Let us now list the propagators that enter in the one loop diagrams for the self-energy of $\chi_3$.
The propagators for the gauge field and Goldstone boson are found to be
\bea
\langle A_\mu^\alpha(p) A_\nu^\beta(-p) \rangle &=& - i \delta^{\alpha \beta}\frac{(p^2 - m_g^2) (g_{\mu\nu}- \frac{p_\mu p_\nu}{p^2 - \xi m_g^2}) + i m_k \epsilon_{\mu\nu\rho}p^\rho+ \xi \frac{p^2-m_k^2-m_g^2}{p^2-\xi m_g^2}p_\mu p _\nu}{(p^2 - m_+^2)(p^2 - m_-^2)}  \nonumber\\
\langle \chi^\alpha(p) \chi^\beta(-p) \rangle &=&  i \frac{\delta^{\alpha \beta}}{p^2 - \xi m_g^2}\,.
\eea
Landau gauge corresponds to the limit $\xi \to 0$ in these expressions. The scalars $(\sigma_\alpha, \t \chi_\alpha)$ have mass mixings, leading to the two-point functions
\bea\label{eq:scalarpropsYMCS}
\langle \sigma_\alpha(p) \sigma_\beta(-p) \rangle &=& i \delta_{\alpha \beta} \frac{p^2-m_g^2}{(p^2-m_+^2)(p^2-m_-^2)} \nonumber\\
\langle \t \chi_\alpha(p) \t \chi_\beta(-p) \rangle &=& i \delta_{\alpha \beta} \frac{p^2-m_g^2-m_k^2}{(p^2-m_+^2)(p^2-m_-^2)} \\
\langle \sigma_\alpha(p) \t \chi_\beta(-p) \rangle &=& -i \epsilon_{\alpha \beta} \frac{m_g m_k}{(p^2-m_+^2)(p^2-m_-^2)} \nonumber\,.
\eea
For the fermions $(\lambda_\alpha, \psi_\alpha)$ we find
\bea\label{eq:fermionpropsYMCS}
\langle \lambda_\alpha(p) \bar \lambda_\beta(p) \rangle &=& i \delta_{\alpha \beta} \frac{(p^2-m_g^2) \not \! p + m_k p^2}{(p^2-m_+^2)(p^2-m_-^2)} \nonumber\\
\langle \psi_\alpha(p) \bar \psi_\beta(p) \rangle &=& i \delta_{\alpha \beta} \frac{(p^2-m_g^2-m_k^2) \not \! p + m_k m_g^2}{(p^2-m_+^2)(p^2-m_-^2)}\\
\langle \lambda_\alpha(p) \bar \psi_\beta(p) \rangle &=& i \epsilon_{\alpha \beta} m_g \frac{ p^2-m_g^2+m_k \not \! p}{(p^2-m_+^2)(p^2-m_-^2)} \nonumber \,.
\eea
Finally, the ghost propagator is
\be
\langle c_\alpha(p) c_\beta(-p) \rangle = i \delta_{\alpha \beta} \frac{1}{p^2 - \xi m_g^2}\,.
\ee
The $i \epsilon$ prescription, not shown here, is the same as in~\cite{Peskin:1995ev}.


\bibliographystyle{JHEP}
\renewcommand{\refname}{Bibliography}
\addcontentsline{toc}{section}{Bibliography}
\providecommand{\href}[2]{#2}\begingroup\raggedright

\end{document}